\documentclass[conference]{IEEEtran}
\IEEEoverridecommandlockouts
\usepackage{cite}
\usepackage{amsmath,amssymb,amsfonts}
\usepackage{algorithmic}
\usepackage{graphicx}
\usepackage{textcomp}
\usepackage{xcolor}
\usepackage{stfloats}
\usepackage{braket}
\usepackage{subcaption}
\usepackage{algorithm}
\usepackage{algorithmic}
\usepackage{booktabs}

\def\BibTeX{{\rm B\kern-.05em{\sc i\kern-.025em b}\kern-.08em
    T\kern-.1667em\lower.7ex\hbox{E}\kern-.125emX}}
\begin{document}

\title{Dependency-Aware Circuit Scheduling for Multi-Core Quantum Systems\\
to Minimize Makespan}
% {\footnotesize \textsuperscript{*}Note: Sub-titles are not captured for https://ieeexplore.ieee.org  and
% should not be used}
% % \thanks{Identify applicable funding agency here. If none, delete this.}
% }

\author{
    \IEEEauthorblockN{
        Rajeswari Suance P S\IEEEauthorrefmark{1},
        Ruchika Gupta\IEEEauthorrefmark{2},
        Maurizio Palesi\IEEEauthorrefmark{3}\IEEEauthorrefmark{1},
        John Jose\IEEEauthorrefmark{1}
    }
    \IEEEauthorblockA{
        \IEEEauthorrefmark{1}Indian Institute of Technology Guwahati, India \quad
        \IEEEauthorrefmark{2}Chandigarh University, India \quad
        \IEEEauthorrefmark{3}University of Catania, Italy \quad
    }
    \IEEEauthorblockA{
        \textnormal{\{s.rajeshwari, johnjose\}@iitg.ac.in, ruchikae7396@cumail.in, maurizio.palesi@unict.it}
    }
}

\maketitle

\begin{abstract}
Multi-core quantum computing architectures have emerged as a promising solution to the qubit scalability limitations of monolithic NISQ devices. Quantum algorithms are expressed as quantum circuits composed of single- and two-qubit gates. However, circuit scheduling in multi-core quantum systems remains largely unexplored. Reducing overall execution time (makespan), increasing core utilization, and hiding communication latency behind computation depends on effective scheduling. In this paper, we first introduce a layered scheduling approach as a baseline where quantum gates within the same layer are executed in parallel, while layers themselves are executed sequentially. We then propose a greedy scheduling strategy which schedules each gate as soon as all its dependencies and required resources are available. This allows fine-grained parallelism across cores. Our evaluation shows that on real benchmarks, greedy scheduling achieves an average 40\% reduction in makespan and improvement in core utilization. The results suggest that the use of intelligent circuit scheduling to exploit parallelism can greatly enhance the speed of circuit execution in multi-core quantum architectures.

\end{abstract}

\begin{IEEEkeywords}
Multi-Core Quantum Computing, Quantum Circuit Scheduling, Network-on-Chip, Quantum Teleportation
\end{IEEEkeywords}

\section{Introduction}

Quantum computing leverages quantum‐mechanical phenomena such as superposition, entanglement, and interference to solve problems that are intractable for classical computers \cite{smith_micro22}. Current noisy intermediate-scale quantum (NISQ) devices are typically fabricated as monolithic chips and face significant scalability challenges \cite{preskill2018quantum}. As the number of qubits increases, issues such as crosstalk, reduced fabrication yield, and increasingly complex control circuitry become more critical \cite{escofet2023interconnect}.

\begin{figure}[t!]  
\centering           
\includegraphics[scale=0.50]{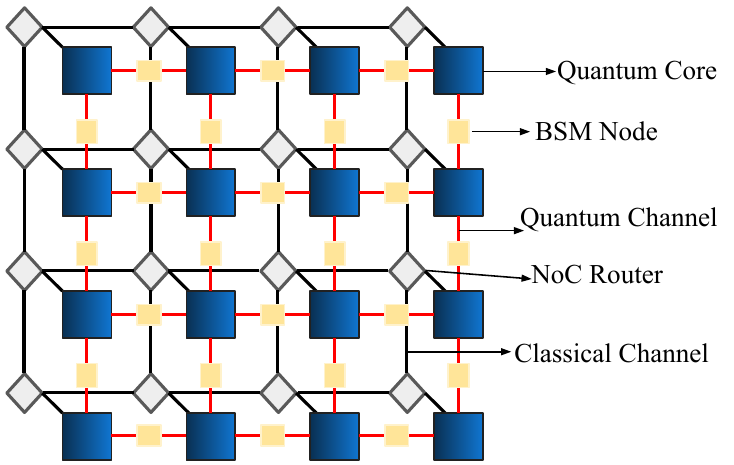}
\caption{Multi-Core Quantum Computing Architecture}
\label{img:architecture}
\end{figure}

To address these limitations, multi-core quantum architectures have emerged as a promising alternative inspired by classical multi-core processors \cite{almudever_date24, smith_micro22, alarcorn_iscas23}. Instead of increasing the number of qubits on a single chip, these architectures distribute qubits across multiple smaller cores, each containing approximately 10–100 qubits. The cores are interconnected through a hybrid quantum-classical Network-on-Chip (NoC) \cite{suance2025decentralized}, enabling scalable and modular system growth while mitigating the physical and control challenges of large monolithic designs. Fig. ~\ref{img:architecture} illustrates such a multi-core quantum computing architecture.

Quantum algorithms are expressed as quantum circuits composed of single- and two-qubit gates. Gates acting on disjoint qubits execute in parallel and are grouped into layers. The total number of such layers is called the depth of the circuit. Two-qubit gates require interacting qubits to be physically adjacent, hence quantum algorithms must undergo circuit compilation to satisfy hardware connectivity constraints \cite{ferrari2023modular, bandic2023mapping, russo2025telesabre}. The final stage of this compilation flow is scheduling, which determines the exact execution time of each gate \cite{vista2025entanglement}. Since qubits are highly sensitive to noise and will rapidly decohere, efficient scheduling is essential to reduce execution time, minimize idle periods, and exploit available parallelism \cite{ferrari2024execution}.

While multi-core architectures improve scalability, they introduce challenges in circuit compilation and scheduling. In these systems, qubits are distributed across cores, and inter-core operations rely on quantum teleportation due to no-cloning theorem \cite{rodrigo2021modelling}. Teleportation-based communication supports mechanisms such as Teledata, which transfers a qubit’s state between cores, and Telegate, which enables distributed gate execution \cite{russo2025telesabre}. However, inter-core communication latency is orders of magnitude higher than local gate execution latency\cite{escofet2023interconnect}, making communication latency a performance bottleneck.

Existing scheduling approaches, particularly in quantum data-center, often rely on strict layered scheduling \cite{vista2025entanglement}. In layered scheduling, gates are grouped into sequential layers that are executed one after another. While this guarantees correctness, it overlooks fine-grained gate dependencies and restricts inter-layer parallelism, leading to longer execution times and underutilized resources. These inefficiencies become even more pronounced in multi-core systems, where communication latency must be carefully overlapped with computation latency to maintain performance.

To address these challenges, we propose a dependency-aware greedy scheduling heuristic that schedules each gate at the earliest possible time when all dependencies are satisfied and required resources including entanglement links are available. By exploiting fine-grained dependencies rather than rigid circuit layers, the proposed method increases parallelism and effectively masks communication latency.

Our major contributions are summarized as follows:
\begin{enumerate}
    \item We formally define the circuit scheduling problem in multi-core quantum architectures, introducing a novel entanglement link-availability constraint specific to NoC-based systems.
    \item We develop an efficient dependency extraction algorithm based on qubit-level tracking.
    \item We propose greedy scheduling algorithm that maximizes parallelism by exploiting fine-grained gate dependencies.
    \item We provide a comparative evaluation of layered and greedy scheduling approaches, demonstrating that the proposed method achieves substantial performance improvements.
\end{enumerate}

The remainder of the paper is organized as follows: Section II presents related works and motivation, Section III describes the proposed scheduling algorithms, Section IV discusses experimental results, and Section V concludes the paper.

\section{Related Work and Motivation}

Current research in multi-core quantum architectures has focused heavily on hardware interconnects and qubit mapping \cite{rodrigo2021modelling, bandic2023mapping, russo2025telesabre}. However, circuit scheduling remains largely unexplored. Existing literature primarily addresses scheduling multiple circuits in cloud environments to improve throughput \cite{ferrari2024execution} or managing centralized entanglement generators \cite{vista2025entanglement}. While standard tools like IBM's Qiskit include circuit scheduling, they target only monolithic architectures \cite{javadi2024quantum}. To the best of our knowledge, this is the first work to address circuit scheduling within multi-core quantum architectures. We propose a strategy that shifts from depth-based metrics \cite{tremba2025circuit} to a dependency-aware approach to minimize makespan.

The primary bottleneck in multi-core systems is the variance in communication latency. If a layer contains both \textit{short-path} (e.g., 1-hop) and \textit{long-path} (e.g., 4-hop) inter-core communications, the entire layer's completion is dictated by the slowest path. This leaves cores involved in local or short-path operations idle, wasting both computational and communicational qubits. Furthermore, precise scheduling is required to extract accurate inter-core network traces, a well-established need in classical NoC research \cite{hestness2010netrace}, that is yet to be addressed for hybrid quantum-classical NoCs. Our work addresses these gaps by shifting from layer-based to dependency-aware greedy circuit scheduling.

% Greedy Scheduling Algorithm
\begin{algorithm}
\caption{Greedy Scheduling}
\label{algo:greedy_scheduling}
\begin{algorithmic}[1]

\STATE \textbf{Input:} Gates $G$, dependencies, mesh topology
\STATE \textbf{Output:} Schedule with $(start, end)$ times

\STATE $est[g] \gets 0$, $eft[g] \gets 0$ for all $g$
\STATE $link\_free[(c_i, c_j)] \gets 0$ for all links

\STATE $ready\_queue \gets \{ g : indegree[g] = 0 \}$
\WHILE{$ready\_queue$ is not empty}
    \STATE $g \gets$ dequeue$(ready\_queue)$
    \STATE $est[g] = \max(eft[p]$ for $p \in predecessors(g))$
    \STATE $gate\_start = est[g]$
    \IF {$g$ spans multiple cores $c_1, c_2$ then}
        \STATE $H \gets \text{compute\_hops}(c_1, c_2)$t
        \STATE $com\_time = H \times \lambda_{hop}$
        \STATE $com\_strt = \max(est[g], link\_free[(c_1, c_2)])$
        \STATE $com\_end = com\_strt + com\_time$
        \STATE $link\_free[(c_1,c_2)] \gets com\_end$
        \STATE $gate\_start \gets com\_end$
    \ENDIF

    \STATE $eft[g] \gets gate\_start + g.duration$
    \STATE \text {Update successors of $g$ and enqueue if ready}

\ENDWHILE

\STATE \textbf{return} schedule

\end{algorithmic}
\end{algorithm}
% Greedy Scheduling Algorithm

\section{Methodology}

\subsection{Multi-Core Architectural Model}
\label{arch_model}
We target a scalable decentralized multi-core architecture \cite{suance2025decentralized}, as shown in Fig. ~\ref{img:architecture}. The system comprises of $M$ cores arranged in an $n \times n$ $2$D mesh topology. Cores are interconnected through Bell State Measurement (BSM) nodes and a hybrid quantum-classical NoC, supporting the entanglement generation required for inter-core qubit teleportation.

Each core contains $Q_i$ qubits. A subset of $k$ qubits serves as communication qubits to establish entanglement links, the remaining $Q_i - k$ are computation qubits. Communication qubits are shared resources managed via FIFO arbitration. Inter-core communication requires hop-by-hop teleportation [4]. The latency for teleporting a state across $H$ hops is $T_{comm} = H \times \lambda_{hop}$, where $\lambda_{hop}$ is the time to generate entanglement and transmit classical data over one hop.

\subsection{Problem Formulation}
Given a quantum circuit with a set of gates $G = \{g_1, g_2, ..., g_{|G|}\}$ operating on $N$ qubits distributed across the $2D$ mesh. The circuit consists of a total of $|G|$ quantum gates. Let $s_g$ and $f_g$ denote the start and finish times of gate $g$ (including inter-core communication if needed). Our objective is to minimize the makespan:

\textbf{Minimize:} 
\[
\text{Makespan} = \max_{g \in G} f_g
\]

\textbf{Subject to:}
\begin{enumerate}
    \item Dependency Constraint: $s_{g_j} \ge f_{g_i} \forall (g_i \rightarrow g_j) \in \mathcal{D}$, where $\mathcal{D}$ is the dependency set.
    \item Entanglement Link Constraint: Each entangled link supports at most one teleportation at a time.
    \item Qubit Constraint: A physical qubit $q$ can only participate in one operation (gate or teleportation) at any given time.
\end{enumerate}

% Example
\begin{figure*}[ht]
    \centering
    % Subfigure (a)
    \begin{minipage}[t]{0.23\textwidth}
        \centering
        \subcaption{$3 \times 3$ Mesh Topology}
        \includegraphics[width=1in, height=0.9in]{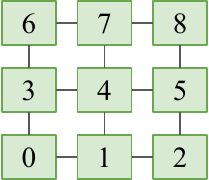}
        \label{img:example_topology}
    \end{minipage}\hfill
    % Subfigure (b)
    \begin{minipage}[t]{0.2\textwidth}
        \centering
        \subcaption{Example Circuit}
        \includegraphics[width=1.2in, height=1in]{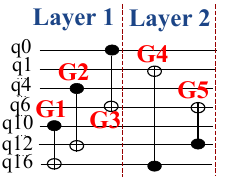}
        \label{img:example_circuit}
    \end{minipage}\hfill
    % Subfigure (c)
    \begin{minipage}[t]{0.25\textwidth}
        \centering
        \subcaption{Gate Scheduling}
        \scriptsize
        \renewcommand{\arraystretch}{0.9}
        \begin{tabular}{c|c|c|c|c}
            Gate & Logical & Core & Hops \\
            & Qubits & Locations & \\
            \midrule
            G1 & (10,16) & (5,8) & 1 \\
            G2 & (4,12) & (2,6) & 4 \\
            G3 & (0,6) & (0,3) & 1 \\
            G4 & (16,1) & (8,0) & 4 \\
            G5 & (12,6) & (6,3) & 1 \\
            \bottomrule
        \end{tabular}
        \label{table_example}
    \end{minipage}\hfill
    % Subfigure (d)
    \begin{minipage}[t]{0.28\textwidth}
        \centering
        \subcaption{Gate Dependency Graph}
        \includegraphics[width=2.25in, height=1in]{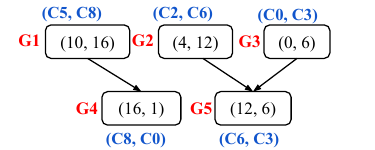}
        \label{img_dag}
    \end{minipage}
    
    \caption{Case study setup (a) $3 \times 3$ Mesh Topology. (b) Example quantum circuit with five two-qubit gates. (c) Table showing gate number, input qubits, their resident cores, and hop counts. (d) Dependency graph of the circuit gates.}
    \label{fig:complete-system}
    
\end{figure*}

\subsection{Circuit Scheduling on Multi-Core Quantum Architectures}

\subsubsection{Layered Scheduling}
Layered scheduling groups gates operating on disjoint qubits into parallel layers \cite{vista2025entanglement}. A strict sequential requirement ensures layer $L_{i+1}$ begins only after $L_i$ fully completes. While this ensures correctness, it limits inter-layer parallelism. If a two-qubit gate in $L_i$ spans distant cores, it delays the entire system, leaving other physical qubits idle and susceptible to decoherence.

\subsubsection{Greedy Scheduling}

To overcome the limitations of layered scheduling, we propose a dependency-aware greedy algorithm, as given in Algorithm ~\ref{algo:greedy_scheduling}. First, we extract directed acyclic graph of gate dependencies by tracking the last gate applied to each logical qubit. Gates with no unresolved dependencies are placed in a ready queue. Each gate is scheduled at its earliest start time ($est$), considering predecessor completion. For multi-core gates, inter-core communication delays are dynamically calculated by checking the availability of entanglement links. Once a gate is scheduled, its successors' in-degrees are decremented, and newly ready gates are queued. This maximizes resource overlap and hides latency.

\subsubsection{Case Study Illustration}
To illustrate the scheduling challenges in a multi-core quantum system, consider a $3\times3$ mesh architecture containing 9 cores, as shown in Fig. ~\ref{img:example_topology}. Assume each core holds 2 physical qubits, giving the system a total of 18 physical qubits. Consider a quantum circuit that requires seven logical qubits and five two-qubit gates, as shown in Fig.~\ref{img:example_circuit}. Each logical qubit $L_i$ is mapped to core $C_i = \lfloor i/2 \rfloor$. Two-qubit gates require inter-core communication, if interacting qubits reside on different cores. This communication is realized through hop-by-hop teleportation \cite{suance2025decentralized}, which employs deterministic XY routing to determine the path and number of hops between source and destination cores. For example, gates $G1$, $G3$, and $G5$ operate on adjacent cores and therefore require only 1 hop. In contrast, gates $G2$ and $G4$ must connect distant cores across the mesh and require four hops. Table~\ref{table_example} summarizes the input logical qubits of each gate, their mapped core locations, and the hops required to transfer qubits for execution. In this case study, each two-qubit gate requires 2 time units to execute, and each hop adds a 1-unit communication delay.

In addition to communication costs, gate execution must respect qubit dependencies: a gate can begin only after all preceding gates acting on its required logical qubits are completed. The gate dependency graph captures these constraints, as shown in Fig. ~\ref{img_dag}. For instance, $G4$ must wait until $G1$ finishes execution on logical qubit $16$, while $G5$ depends on the completion of both $G2$ on qubit $12$ and $G3$ on qubit $6$. These communication delays and dependencies directly impact scheduling efficiency.

\begin{figure}[t!]
\centerline{\includegraphics[width=\columnwidth]{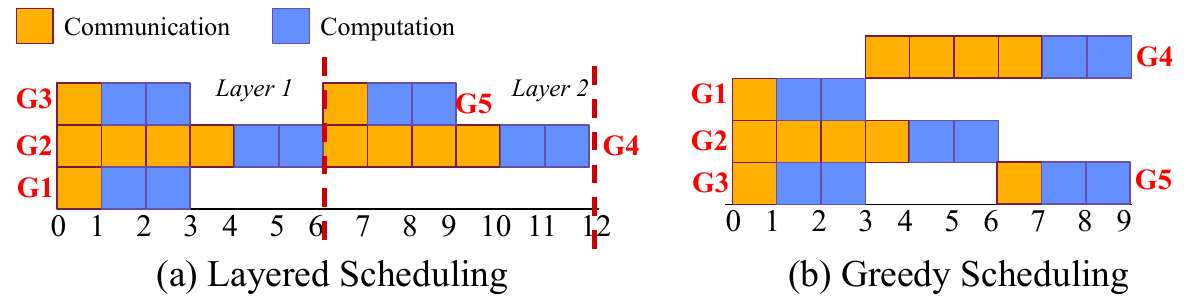}}
\caption{Comparison of scheduling strategies: layered scheduling (makespan=12) versus greedy scheduling (makespan=9)}
\label{scheduling_example}
\end{figure}

Under layered scheduling, gates are executed strictly layer by layer. Layer 1 contains both a short-latency 1-hop gate ($G1$) and a long-latency 4-hop gate ($G2$). While $G1$ completes in 3 time units, $G2$ requires 6 time units. Because layered scheduling enforces rigid layer boundaries, layer 2 cannot begin until the slowest gate in layer 1 finishes. Consequently, the system remains idle after $G1$ completes and must wait until $T=6$. Since layer 2 also includes a 4-hop gate ($G4$), it requires another 6 time units, resulting in a total makespan of $T=12$, as shown in Fig.~\ref{scheduling_example}a. 

In contrast, the greedy scheduling approach removes this bottleneck by exploiting dependency awareness. Since $G4$ depends only on the early-completing $G1$, it can be scheduled immediately after $G1$ finishes at $T=3$, without waiting for $G2$. As a result, $G4$ executes concurrently with the long-latency $G2$, creating inter-layer parallelism. This overlap effectively hides the communication delay of $G4$, reducing the overall makespan to $T=9$, as illustrated in Fig.~\ref{scheduling_example}b.

% Synethtic Benchmark
\begin{figure*}[ht!]
    \centering
    \begin{subfigure}[t]{0.30\textwidth}
        \centering
        \includegraphics[width=50mm, scale=0.75]{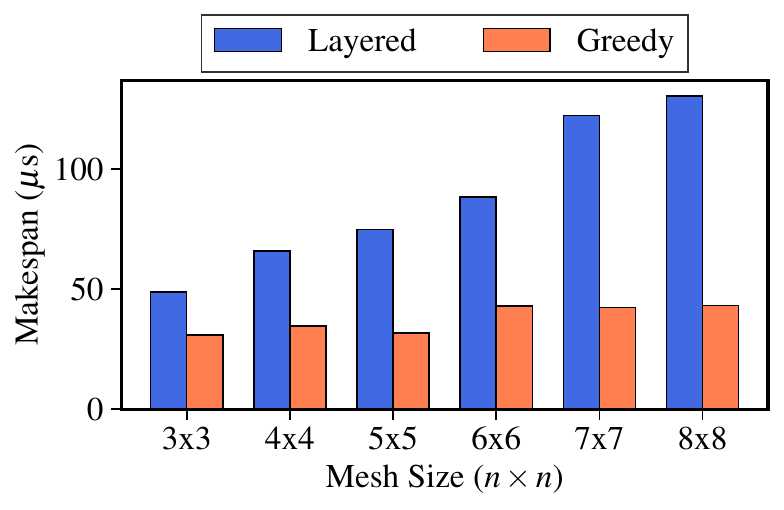}
        \caption{Makespan}
        \label{synth_makespan}
    \end{subfigure}
    \hspace{0.03\textwidth} 
    \begin{subfigure}[t]{0.30\textwidth}
        \centering
        \includegraphics[width=50mm, scale=0.75]{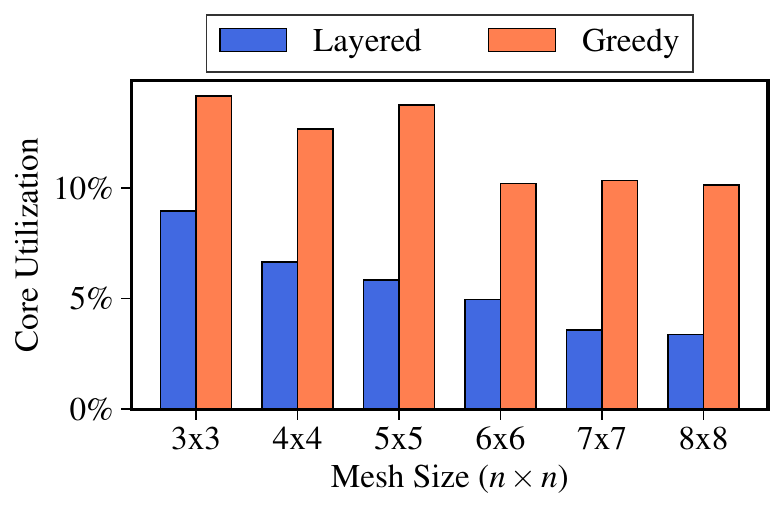}
        \caption{Core Utilization}
        \label{synth_util}
    \end{subfigure}
    \hspace{0.03\textwidth} 
    \begin{subfigure}[t]{0.30\textwidth}
        \centering
        \includegraphics[width=50mm, scale=0.75]{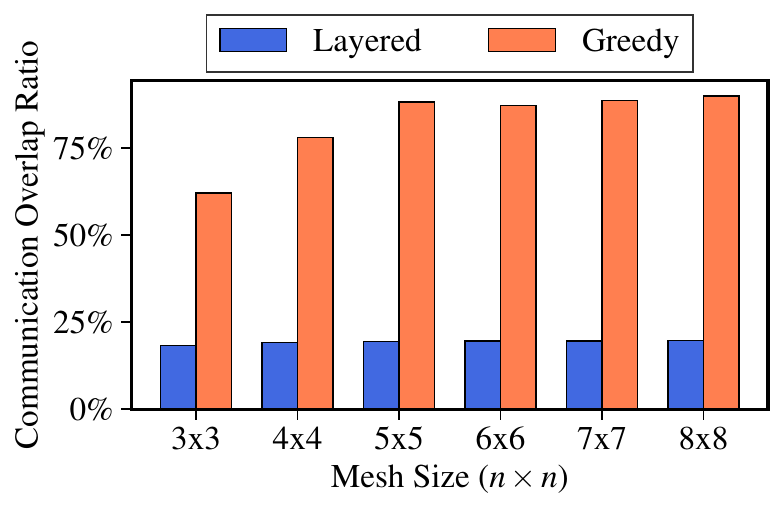}
        \caption{Communication Overlap Ratio}
        \label{synth_comm}
    \end{subfigure}
    \caption{Comparison of various circuit scheduling with synthetic random circuits}
    \label{fig:synt}
\end{figure*}
% Synethtic Benchmark

\section{Evaluation}
\subsection{Experimental Setup}

As discussed in Section ~\ref{arch_model}, we consider a decentralized multi-core architecture with an $n \times n$  mesh topology, where the total number of cores is $M$ \cite{suance2025decentralized}. The system consists of $N$ qubits distributed across $M$ cores, each containing $Q_i$ qubits. Of these, $k$ qubits serve as communication qubits for entanglement, and the remaining $Q_i - k$ are computation qubits. We set $Q_i = 32$ and $k = 4$, since each core has four output directions, this allocates one communication qubit per direction. This allows only one entanglement link between adjacent cores at a time, ensuring inter-core transfers occur sequentially between a core and its direct neighbors.

To evaluate the circuit scheduling algorithms, we adopt a strong-scaling approach. In this setting, the number of qubits per core, $Q_i$, is fixed while the number of cores $M$ and the circuit size are increased. This method enables us to assess communication overhead and its impact on computational performance. For our experiments, we vary the number of cores by increasing the mesh size from $3 \times 3$ to $8 \times 8$. 

We take as input a quantum circuit generated using Qiskit \cite{javadi2024quantum} or MQT benchmark circuits \cite{quetschlich2023mqt}. The focus of this work is specifically on optimizing the circuit scheduling stage. For simplicity, we adopt a trivial initial qubit mapping.

We assume the following gate execution times, based on a superconducting-qubit architecture: single-qubit gates require $5$ \textit{ns} and two-qubit gates require $50$ \textit{ns} \cite{bahrani2023analysing}. For teleportation, we assume guaranteed entanglement generation at a fixed rate \cite{vista2025entanglement}. We use a $10$ \textit{MHz} entanglement generation rate and a classical-communication latency of $150$ \textit{ns} \cite{hashim2025efficient}. Hardware noise is not considered in this work and is left for future study.

\textbf{Benchmarks} We create synthetic benchmarks by generating random circuits using Qiskit \cite{javadi2024quantum} with circuit depth fixed at 10 for all experiments. For real benchmarks, we use circuits from the MQT benchmark suite \cite{quetschlich2023mqt}. Specifically, we evaluate four circuits: Graph State preparation, Portfolio Optimization with Quantum Approximate Optimization Algorithm (QAOA), Quantum Fourier Transform (QFT), and Quantum Phase Estimation (QPE). For both synthetic and real benchmarks, we define circuit size as the number of qubits available for computation, excluding qubits reserved for communication.

\subsection{Performance Metrics}
We evaluate our scheduling approach using three metrics:
\begin{enumerate}
\item \textbf{Makespan ($T_{\text{total}}$):} The total circuit execution time defined as the maximum earliest finishing time ($\mathrm{eft}$) across all gates $g$. $T_{\text{total}}=\max_{g}\mathrm{eft}[g]$.
\item \textbf{Core Utilization:} The percentage of time $N$ cores actively execute gates (excluding teleportation latency). It is computed as $\frac{\sum_{c}T_{\text{busy}}(c)}{N\times T_{\text{total}}}\times 100$, where $T_{\text{busy}}(c)$ is the active gate execution time on core $c$.
\item \textbf{Communication Overlap Ratio ($R_{\text{comm}}$):} The fraction of total communication time ($T_{\text{comm}}$) that is successfully hidden behind parallel gate execution ($T_{\text{overlap}}$), defined as $R_{\text{comm}}=\frac{T_{\text{overlap}}}{T_{\text{comm}}}$.
\end{enumerate}

\subsection{Results and Discussion}

Fig. ~\ref{synth_makespan} shows the comparative study of makespan for synthetic random circuits across different hardware configurations. We can see that proposed greedy scheduling algorithm consistently achieves a lower makespan than layered scheduling. This improvement arises because the greedy approach exploits inter-layer parallelism, allowing gates to execute as soon as their dependencies and resources are available. In contrast, the layered scheduling ignores the fine-grained parallelism and fails to hide inter-core communication latency, which is particularly evident for the larger meshes.

Fig. ~\ref{synth_util} shows the comparison of core utilization for the synthetic benchmarks. We can see that the greedy scheduling algorithm improves utilization across different mesh configurations. However, core utilization does not exceed 15\% for either scheduling algorithm. This is expected because only a single circuit is distributed across all cores, generating a large amount of inter-core communication relative to gate execution. If multiple circuits were executed concurrently, computation would dominate communication, leading to significantly higher utilization. We leave this multi-circuit scenario for future study.

Fig. ~\ref{synth_comm} shows the percentage of inter-core communication that overlaps with computation across mesh configurations. Layered scheduling achieves at most $20\%$ overlap for all mesh sizes. In contrast, greedy scheduling significantly increases the overlap, reaching up to $76\%$ and saturating beyond the $6 \times 6$ mesh configuration. This indicates that greedy scheduling is more effective at hiding communication latency behind gate execution, thereby reducing overall execution time.
\begin{figure*}[t!]
    \centering
    \begin{subfigure}[t]{0.3\textwidth}
        \centering
        \includegraphics[width=50mm, scale=0.4]{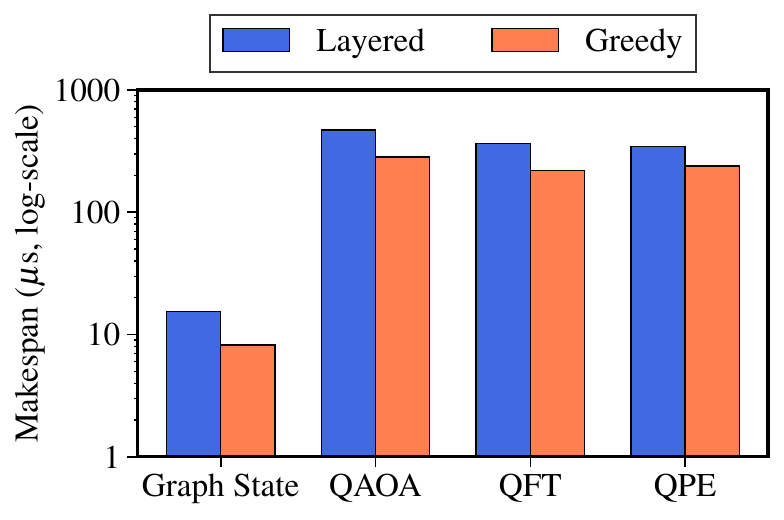}
        \caption{Makespan}
        \label{real_makespan}
    \end{subfigure}
    \hspace{0.03\textwidth} 
    \begin{subfigure}[t]{0.30\textwidth}
        \centering
        \includegraphics[width=50mm, scale=0.75]{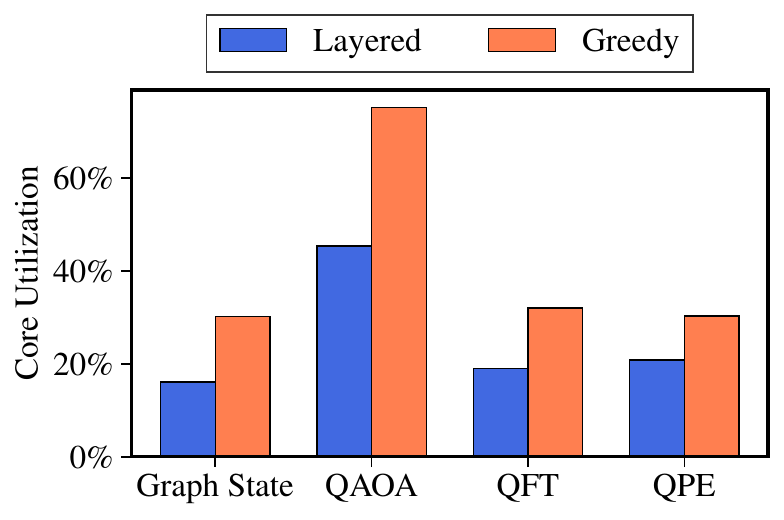}
        \caption{Core Utilization}
        \label{real_util}
    \end{subfigure}
    \hspace{0.03\textwidth} 
    \begin{subfigure}[t]{0.30\textwidth}
        \centering
        \includegraphics[width=50mm, scale=0.75]{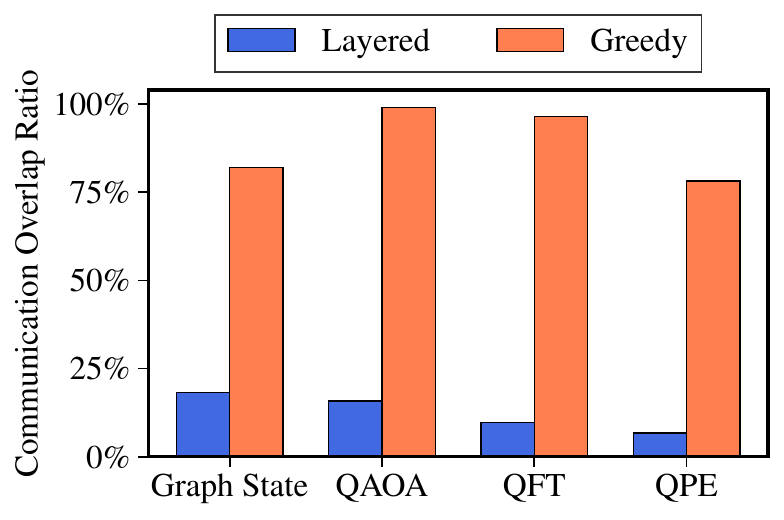}
        \caption{Communication Overlap Ratio}
        \label{real_comm}
    \end{subfigure}
    \caption{Comparison of various circuit scheduling results with real benchmarks on a  $4 \times 4$ mesh with 32 qubits per core}
    \label{fig:main}
\end{figure*}

To validate the correctness of the scheduling algorithms, we evaluate real circuits from the MQT benchmark suite on a $4 \times 4$ mesh with 32 qubits per core. Fig.~\ref{real_makespan}, \ref{real_util}, and \ref{real_comm} show that greedy scheduling consistently outperforms layered scheduling across all benchmarks, achieving an average $40\%$ reduction in makespan, along with higher core utilization and improved communication–computation overlap. These results confirm the overall effectiveness of greedy scheduling.

A key limitation of this study is the assumption of deterministic teleportation with always successful entanglement generation. A more realistic scenario can be explored by generating traffic from the schedule to test NoC communication protocols. This approach would also allow the investigation of contention on entanglement links when multiple requests compete for the same output direction, which remains an open challenge for future work.

\section{Conclusion}
Our study in this paper addressed a key problem in multi-core quantum computing: scheduling circuits while balancing communication overhead and parallel execution. We first implemented layered scheduling as a baseline and then enhanced it with a greedy strategy that scheduled gates as soon as resources and dependencies became available. Experiments on both synthetic and real benchmarks demonstrated clear benefits, including shorter makespan, higher core utilization, and greater overlap between communication and computation time. These results highlighted the importance of exploiting fine-grained parallelism to accelerate quantum circuit execution and reduce the risk of qubit decoherence.

\section*{Acknowledgment}
The authors gratefully acknowledge funding from the European Commission through HORIZON-EIC-2022-PATHFINDEROPEN01-101099697 (QUADRATURE).

% \section*{Acknowledgment}

% The preferred spelling of the word ``acknowledgment'' in America is without 
% an ``e'' after the ``g''. Avoid the stilted expression ``one of us (R. B. 
% G.) thanks $\ldots$''. Instead, try ``R. B. G. thanks$\ldots$''. Put sponsor 
% acknowledgments in the unnumbered footnote on the first page.

\bibliographystyle{IEEEtran}
\bibliography{reference}

\end{document}